\documentclass[12pt,a4paper]{article}
\usepackage[utf8]{inputenc}
\usepackage{amsmath}
\usepackage{amsfonts}
\usepackage{amssymb}
\usepackage{graphicx}

\usepackage{bm}
\usepackage{url}

\usepackage{natbib}
\setcitestyle{authoryear,open={(},close={)}}

\title{A modelling approach for correcting reporting delays in disease surveillance data}

\author{Leonardo S Bastos\footnote{Scientific Computing Program, Oswaldo Cruz Foundation, Rio de Janeiro, 21040-360, Brazil}
\and
Theodoros Economou\footnote{Department of Mathematics, University of Exeter, Exeter, EX4 4QF, UK} 
\and
Marcelo F~C Gomes \footnotemark[1]
\and
Daniel A~M Villela \footnotemark[1]
\and
Flavio C Coelho \footnote{School of Applied Mathematics, Getulio Vargas Foundation, Rio de Janeiro, 22250-900, Brazil}
Oswaldo G Cruz \footnotemark[1]
\and
Oliver Stoner  \footnotemark[2]
\and
Trevor Bailey  \footnotemark[2]
\and
Claudia T Code\c{c}o \footnotemark[1]}

\date{Published in Stats in Medicine, 2019. DOI: \url{ https://doi.org/10.1002/sim.8303}}

\begin{document}

\maketitle

\begin{abstract}
One difficulty for real-time tracking of epidemics is related to reporting delay. The reporting delay may be due to laboratory confirmation, logistical problems, infrastructure difficulties and so on. The ability to  correct the available information as quickly as possible is crucial, in terms of decision making such as issuing warnings to the public and local authorities. A Bayesian hierarchical modelling approach is proposed as a flexible way of correcting the reporting delays and to quantify the associated uncertainty. Implementation of the model is fast, due to the use of the integrated nested Laplace approximation (INLA). The approach is illustrated on dengue fever incidence data in Rio de Janeiro, and Severe Acute Respiratory Infection (SARI) data in Paran\'a state, Brazil.

\end{abstract}

\section{Introduction}
\label{sec:intro}
Surveillance systems play a crucial role in managing infectious disease risk. The main requirements for a {good surveillance system} are timeliness, sensitivity and  specificity, together with readily interpretable outputs \citep{farrington1996statistical}. Timeliness reflects the speed or delay between steps in a surveillance system \citep{CDCguideline}: the time between the onset of an adverse health event and its report, and the time between report and the identification of trends or outbreaks, for example.

Disease surveillance in most countries is passive, relying on the cases reported by health care providers from patients seeking care. The number of cases reported quite commonly suffers a reporting delay that can vary across localities, being susceptible to the adherence of local health care providers to the reporting protocol, as well as the access of patients to health care. Timeliness is also affected by conflicting factors due to the disease incidence: delays may decrease during the high transmission season because of awareness among doctors and patients; conversely, delays may increase during high transmission seasons due to the saturation of the health care system. {Reporting delays, especially ones whose structure varies in time, distort the relationship between the reported disease incidence and the true disease incidence. Surveillance and warning systems relying on reported incidence to assess risk can therefore be misinformed, if this delay is not somehow corrected.}

From a statistical perspective, reporting delay is a censoring problem, albeit one for which the {observable (reported)} data will eventually become available. {Note that we make a distinction here between observable data and the truth. We term observable those incidence cases that were detected and eventually reported. In disease surveillance however, data are always potentially under-reported, i.e. disease cases that were never detected or that were detected but never reported. As such, the true disease count is the observable count plus any cases that were {never reported}. In this paper, we focus on correcting reporting delay in the observable data, noting that correcting for under-reporting is generally a non-trivial task, requiring additional sources of information such as prior knowledge on under-reporting rates or a sample of fully observed data (i.e. the truth) as discussed in \cite{Stoner2019}.}

The aim of this paper is to propose a flexible statistical modelling framework that enables the estimation of the missing {(observable)} data in order to perform nowcasting as well as the potential for forecasting. The framework was developed with two goals in mind: to be a useful decision making tool while at the same time being flexible enough to apply to a range of problems with complex data structures and to provide reliable corrections as well as full quantification of uncertainty. To achieve these goals, the framework should possess the following attributes:
\begin{itemize}
 \item 
 Practical (computational) feasibility. This is vital if the model is to be used in conjunction with a warning system which can be potentially updated in real-time.
 \item
 Flexibility. The model should readily allow for covariates relating to the delay mechanism, the variability of the disease, as well as other relevant information (such as Twitter feeds and weather nowcasts/forecasts).
 \item
 Complexity. The model should be able to capture any (residual) spatio-temporal variability, both in the delay mechanism and the progression of the disease. Temporal dependence is particularly important in being able to detect outbreaks.
\end{itemize}

Furthermore, with this being a prediction problem, a Bayesian formulation is desirable as it enables use of predictive distributions that quantify all the associated uncertainty in correcting the missing values. The motivation behind such a modelling framework is presented in the model application section of this paper, where the overall goal is to develop a real-time online warning system for infectious diseases in Brazil. The model is used to correct the number of dengue cases in Rio de Janeiro and also the number of Severe Acute Respiratory Infection (SARI) cases in Paran\'{a} state in Brazil.

The paper is structured as follows. In Section~\ref{sec:background}, we present the formulation of the problem, set the relevant notation and discuss current approaches to model reporting delay. In Section~\ref{sec:methods}, we present the modelling framework and how to perform inference to obtain the predictive distribution of the reporting cases. In Section~\ref{sec:application}, we apply the model to dengue data from Rio de Janeiro and to Severe Acute Respiratory Illness (SARI) data from the state of Paran\'{a}, Brazil. Finally, in Section~\ref{sec:discussion} we provide a summary and discussion of the results obtained.

\section{Background}\label{sec:background}
\subsection{The run-off triangle}
\begin{figure}[ht!] 
\centering
\includegraphics[width=0.8\textwidth]{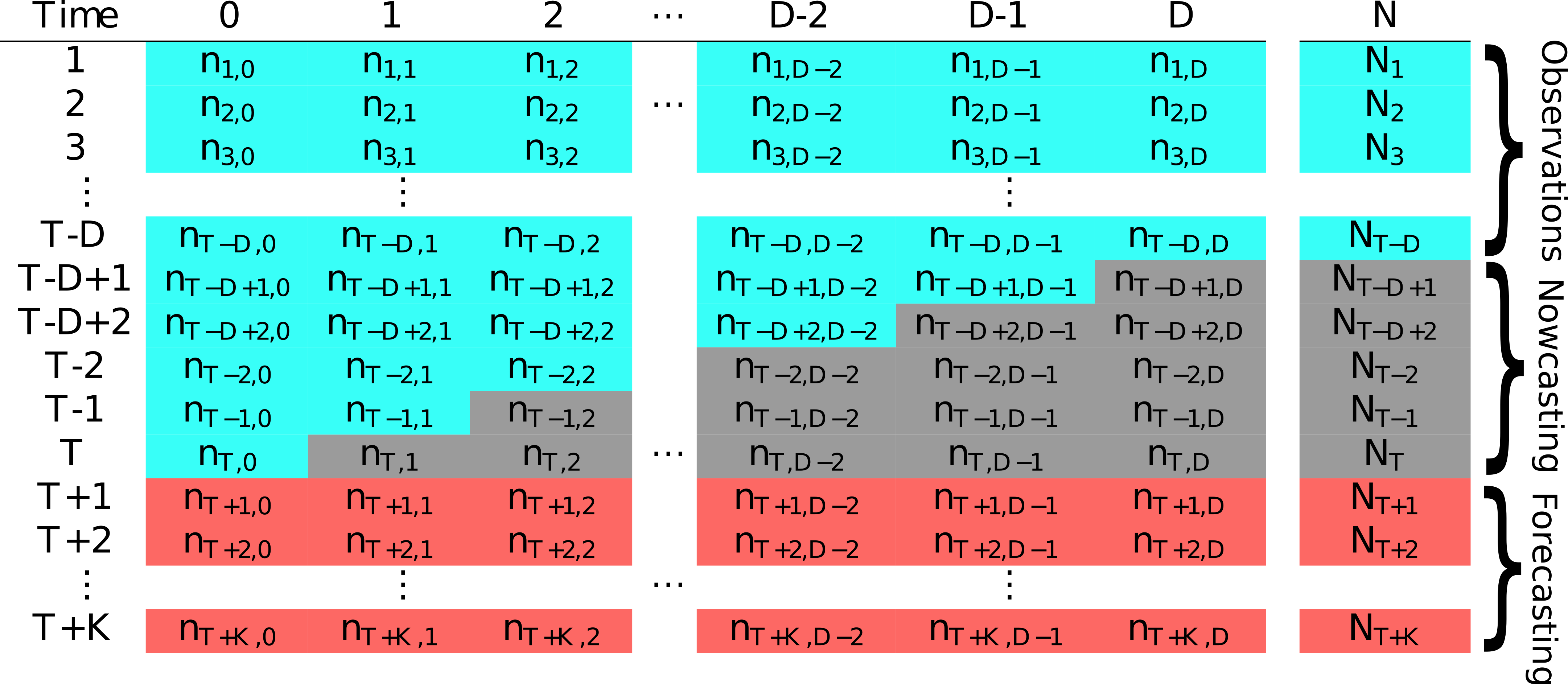}
\caption{Table illustrating the typical data structure in a reporting delay problem. The values in the blue cells are fully observed number of cases at time as of time $T$ (today); the values in grey are the occurred-but-not-yet-reported number of events (run-off triangle); and the values in red are the future number of event we may be interested to forecast.} 
\label{table1}
\end{figure}

The typical data structure for the reporting delay problem is given in Figure \ref{table1} where the rows correspond to time $t=1,2,\ldots,T$ and the columns correspond to amount of delay (in the same units as $t$), $d=0,1,\ldots,D$, where $D$ is the maximum possible delay. For any time step (row) the true total amount of events (e.g. disease occurrences) is $N_t=\sum_{d=0}^Dn_{t,d}$ so that $n_{t,d}$ is the number of events that occurred at time $t$, that were reported at $d$ time steps after $t$ (with $n_{t,0}$ being the number that were actually reported at $t$). Assuming for simplicity that $T$ is `today', then the values $n_{t,d}$ in the grey boxes of Figure \ref{table1} are missing and so are the corresponding the totals $N_t$. These occurred-but-not-yet-reported events are also called the run-off triangle \citep{Mack1993}, all values of which potentially need to be estimated for accurate risk assessment (e.g. for detecting a sharp increase in occurrences).

For reliable risk assessment at time point $T$, the counts in the run-off triangle need to be estimated (nowcast), ideally along with the uncertainty associated with doing so. The following section discusses some recent approaches to this problem, along with the motivation for the one proposed in this paper.

\subsection{Current approaches}
The problem of reporting delay is not unique to epidemiological data. It has also been identified in actuarial science where there may be delay between insured damage and the associated insurance claim, so that the challenge is to estimate the number of outstanding claims \citep{Renshaw1998}. Broadly speaking, two modelling frameworks have been developed to tackle the reporting delay problem.

The first approach is to consider the distribution of the counts $n_{t,d}$ conditional on the totals $N_t$. The framework is then hierarchical where the $N_t$ are assumed to be distributed as Poisson or Negative Binomial, and then $n_{t,k}|N_t$ is Multinomial with some probability vector of size $D$ that needs to be estimated. 
This framework was used in a Bayesian nowcasting model to correct delays in the reporting of Shiga toxin-producing Escherichia coli in Germany \cite{Hohle2014}. 
The model allows for smooth changes in the temporal variation of the total number of cases $N_t$ as well as in the delay mechanism by characterising the Multinomial probability vector as a function of time. Furthermore, a test for detecting outbreaks in infectious disease on the basis of this conditional approach have been developed \cite{Noufaily2016}.

The other approach, primarily utilised in correcting insurance claims, is to think about the distribution of the cell counts $n_{t,d}$ directly. The so called chain-ladder technique was developed as a distribution-free method to estimate the missing delayed counts \citep{Mack1993}. Later it was shown that the underlying model for the chain-ladder technique is a Generalized Linear Model for $n_{t,d}$ where the mean is characterised as $\mathbb{E}[n_{t,d}] = \lambda_{t,d} = \mu + \alpha_t + \beta_d$ \cite{Renshaw1998}. The model has been extended in many ways to accommodate for various parametric and non-parametric functional forms as well as potential covariates in $\lambda_{t,d}$, see for instance \cite{England2002, Barbosa2002}. {It is interesting to note, that the chain ladder framework can be motivated from the conditional Multinomial approach as was shown in \cite{Salmon2015}. Assume first that the total counts $N_t$ arise from a Negative Binomial distribution with some mean $\lambda_t$ and dispersion parameter $\phi$. This is a common assumption when modelling disease count data, where the Negative Binomial extends the Poisson to allow for overdispersion in data where the amount of susceptible population is not actually known--a common problem in observational surveillance data \citep{Held2005}. Second, assuming the counts in each row of Table \ref{table1} are conditionally Multinomial, $\bm{n}_t \sim MN(\bm{\pi}_t,N_t)$, then it can be shown that the marginal distribution of each $n_{t,d}$ is a Negative Binomial with mean $\pi_{t,d}\lambda_t$ and dispersion parameter $\phi$. In this way, the chain ladder method which directly models the marginals as Negative Binomial can be justified from the conditional Multinomial approach (noting however that $\pi_{t,d}$ and $\lambda_t$ can not be separated).}

{Here we extend the chain ladder approach with Negative Binomial marginals, to allow for spatio-temporal variation} in the counts as well as covariate effects. Spatial variation is something that has not yet been considered in the various approaches to date, however it is important to appreciate that both the delay mechanism as well as the temporal variability in the process giving rise to the counts can vary is space. For instance, in the application of the model in section \ref{sec:sari}, the delay mechanism in reporting of Severe Acute Respiratory Infection (SARI) in Brazil is allowed to vary in space to account for the differences in the reporting process across administrative regions and to also borrow information across these regions. Furthermore, the particular formulation of the model that we propose, readily allows for dependence along both the columns and rows of Figure \ref{table1}, to capture the temporal variability of the disease occurrence as well as the temporal structure of the delay mechanism.

{Note that all the approaches mentioned here are purely statistical, in the sense that there is no specific component in the models relating to the disease dynamics. Incorporating mechanistic or physical elements relating to the disease, can greatly improve predictions by allowing the science to effectively inform the modelling. Approaches combining mechanistic models such as SIR (Susceptible-Infectious-Recovered) and statistical ones have been utilised in the past to model disease time series (e.g. \cite{Finkenstadt2000}), however such modelling efforts require well-documented data and can be computationally expensive. As mentioned, the focus here is on observational surveillance studies (often lacking in vital information such as the amount of susceptible population) and efficient modelling for use in real-time decision making.}

\section{Model specification}
\label{sec:methods}

Recall that $n_{t,d}$ is a random variable describing the number of events that occurred at time $t=1,2,\ldots,T$ but not reported until $d=0,1,2,\ldots,D$ time units later. $T$ is the last time step for which data is available, and $D$ is the maximum acceptable delay, {which for disease applications is potentially infinite but for simplicity we assume that $D$ is bounded (this could also be true for insurance claims that must be filed within a certain period of the event).} We model $n_{t,d}$ with a (conditional) Negative Binomial distribution with mean $\lambda_{t,d}$ and scale parameter $\phi$, i.e.
\begin{equation}
\label{likelihood}
n_{t,d} \sim NegBin(\lambda_{t,d}, \phi), \qquad \lambda_{t,d} >0, \quad \phi>0.
\end{equation}
The parameterisation used here is such that $\mathbb{E}[n_{t,d}] = \lambda_{t,d}$ and $\mathbb{V}[n_{t,d}] = \lambda_{t,d} (1 + \lambda_{t,d} / \phi)$. As mentioned in section \ref{sec:intro}, we take a Bayesian approach so that predictive distributions of $n_{t,d}$ for any $t$ and $d$ (given the data) are readily available as well as all the associated uncertainty in their estimation. As the dispersion parameter $\phi$ approaches infinity, the Negative Binomial reduces to the Poisson distribution. As such, $\phi$ can be though of as a parameter that adds variability and thus flexibility to the Poisson. We assume en exponential $\mbox{Exp}(0.1)$ prior distribution for $\phi$ with mean 10 and standard deviation 10. This is a weakly informative prior which places more probability over smaller values of $\phi$ and thus assumes preference of the Negative Binomial to the Poisson. {The term ``weakly informative prior'' is used here to emphasize that the prior is not an ellicited infomative prior, nor a completely vague ``infinite'' variance prior.} 

{To capture structured temporal variability in $n_{t,d}$, the logarithm of their mean, $\lambda_{t,d}$, is characterized as follows:
\begin{equation} \label{lambda}
\log(\lambda_{t,d}) = \mu + \alpha_t + \beta_d + \gamma_{t,d} + \eta_{w(t)} + \bm{X}'_{t,d}\bm{\delta},
\end{equation}
where $\mu$ is the overall mean count at the log-scale and $\bm{X}'_{t,d}$ is a matrix of temporal and delay-related covariates with associated vector of parameters $\bm{\delta}$.} The random effects $\alpha_t$ capture the mean temporal evolution of the count generating process, while the $\beta_d$ capture the mean structure of the delay mechanism. These can be modelled using random walks, in the simplest case a first-order ones, i.e.
\begin{equation} \label{alphat}
\alpha_t  \sim  N( \alpha_{t-1}, \sigma^2_\alpha ), \quad t=2,3,\ldots,T, 
\end{equation}
and 
\begin{equation} \label{betad}
\beta_d  \sim  N( \beta_{d-1}, \sigma^2_\beta), \quad d=1,2,\ldots,D, 
\end{equation}
where half Normal $\mbox{HN}(\tau^2)$ prior distributions are assumed for $\sigma_{\alpha}$ and $\sigma_{\beta}$. These are distributions on $[0,\infty)$ where parameter $\tau$ controls the variance. Thinking about $\alpha_t$ and $\beta_d$ as unknown functions in time and delay, then $\tau$ controls the {"wiggliness"} of these functions---the smaller it is, the {less wiggly (or in some sense `smooth')} the functions will be (i.e. the smaller the first order differences will be). Noting that these random effects influence the mean count at the log-scale, for $\beta_d$, we choose $\tau=1$ while for $\alpha_t$ we choose $\tau=0.1$. These are weakly informative priors, reflecting our belief that first order differences across the columns (delay) will be bigger than first order differences along the rows (time). In other words, the temporal trend is assumed {less wiggly} a-priori than the delay structure, though this assumption can be overridden by data given sufficient evidence. Note also that adding temporal dependence through $\alpha_t$ is common in modelling disease counts \citep{Bauer2016}, and allows for temporal variation in the process giving rise to the counts, other than what may be explained by temporal covariates $\bm{X}'_{t,d}$ such as weather patterns. {In addition, it is worth mentioning that if it is thought that the count of infections has potentially much longer temporal memory, e.g. if the infectious and incubation periods are longer than one time unit, then higher order random walks can be used.}

{The time-delay interaction term $\gamma_{t,d}$ is modelled as
\begin{equation} 
\gamma_{t,d}  \sim  N( \gamma_{t-1,d}, \sigma^2_\gamma )
\end{equation}
so that there is an independent realisation of a random walk order 1, for each delay column. This term is important, as it allows for changes in the delay mechanism over time. This is something we would expect since, for example, it is empirically known that delays are more severe during an outbreak due to prioritisation of treating patients rather than reporting cases. It will also indirectly allow for non-zero correlation across the columns in Figure \ref{table1}, which in turn affect the variance in the totals $N_t$. The prior on $\sigma^2_\gamma$ is the same as the one on $\sigma^2_\alpha$. Lastly, $\eta_{w(t)}$, where $w(t)=1,\ldots,52$ is the week index, is a seasonal component defined as a second order random effect:
\begin{equation} 
\eta_{w} \sim  N( 2\eta_{w-1}-\eta_{w-2}, \sigma^2_\eta)
\end{equation}
constrained in such a way that week 1 and week 52 are joined. This term is also important as it can capture temporal variability in disease incidence that varies with time of year. For mosquito-borne disease (such as ones we model here), the incidence rate is strongly linked to weather variation which in turn is seasonally varying. The variance parameter $\sigma^2_\eta$ is less interpretable than in first order random walks, but in general smaller values will result in a {less wiggly} function. We choose a $\mbox{HN}(1)$ prior to allow enough flexibility while restricting values that are too extreme. All of the components $\alpha_t$, $\beta_d$, $\eta_{w(t)}$ and $\gamma_{t,d}$ are constrained to sum to zero, to allow identifiability of the intercept $\mu$.}

{Figure \ref{delay.fig} shows a time series of weekly dengue occurrences in Rio de Janeiro. This is an archetypal example of data we wish to model, exhibiting periods of very low activity but also sharp increases (outbreaks) as well as decreases. The auto-regressive nature of the temporal random effects has the benefit of being able capture such behaviour in time, utilising the short term memory in the process to adapt as new data become available. Alternative ways of characterising temporal structure such as {conventional} Gaussian process priors may be too smooth to capture this behaviour while models based on penalised splines can suffer from the same issue.} Similarly, the autoregressive effects $\beta_d$ allow for flexible characterisation of the delay mechanism (as illustrated in section \ref{sec:application}.) {Note however that it is very important that the temporal structure is captured adequately using model checking as is performed later in section \ref{sec:application}. Failing that, more flexible random effect distributions can be considered, such as a mixture of Gaussian distributions \citep{faulkner2018}.}

The posterior distribution for $\bm{\Theta} = (\mu, \{\alpha_t\}, \{\beta_d\}, \{\gamma_{t,d}\}, \{\eta_{w}\}, \sigma^2_\alpha, \sigma^2_\beta, \sigma^2_\gamma, \sigma^2_\eta, \phi)$ given all the observed data $\bm{n}=\{n_{t,d}\}$ is given by 
\begin{equation} \label{posterior}
p( \bm{\Theta} \mid \mathbf{n} ) \propto p ( \bm{\Theta} ) \prod_{t=1}^{T} \prod_{d=0}^{D}p(n_{t,d} \mid \bm{\Theta}) 
\end{equation}
where $p(n_{t,d} \mid \bm{\Theta})$ is the Negative Binomial density function \eqref{likelihood}, and $p(\bm{\Theta})$ is the joint prior distribution given by the product of the prior distributions for $\phi$, $\sigma^2_\alpha$, $\sigma^2_\beta$, $\sigma^2_\gamma$, $\sigma^2_\eta$, and the random effects distributions. {A list of all prior distributions used, as well as a simulation experiment checking the plausibility of the data compared to simulation from the prior predictive distribution, are given as supplementary material in the following GitHub page: \url{https://github.com/lsbastos/Delay}.}

\subsection{Model implementation}
Samples from the posterior distribution (\ref{posterior}) can be obtained via traditional Markov chain Monte Carlo (MCMC) methods \citep{gamerman2006markov} using for instance software such as NIMBLE \citep{nimble}. MCMC however, can be computationally intensive especially when the model is extended to allow for spatial variation as discussed in the next subsection. A more efficient approach would be to obtain approximate samples from the posterior distribution of $\bm{\Theta}$ using integrated nested Laplace approximation or INLA \citep{Rue2009} using a copula approach already implemented in the INLA package for R (\url{www.r-inla.org}).

The INLA approach to obtaining samples from the posterior distribution can be significantly faster than MCMC \citep{Rue2017}, with the added benefit of reduced user input (e.g. to assess convergence of the Markov chains). Implementing the proposed model in R-INLA, makes it an attractive decision making tool for correcting reporting delay in real-time, e.g. using an online interface for issuing warnings, as discussed in section \ref{sec:application}. The key concept in INLA is to combine nested Laplace approximations with numerical methods for sparse matrices for efficient implementation of latent Gaussian models. Since the model proposed here is in fact a latent Gaussian model (i.e. the joint distribution of the random effects is multivariate Gaussian) it can be readily implemented using R-INLA.

\subsection{Spatial variation} \label{sec:spatial}

In many applications, including the ones considered in this paper, the data may be spatially grouped, e.g. into a number of administrative regions spanning Brazil. In general, the model presented above can be implemented independently for the various spatial regions/locations. In practice however, it would make more sense to analyse all data together by extending the model to  allow for spatial variation not only in the process giving rise to the counts, but also in the delay mechanism. This allows for pooling of information to aid estimation in spatial locations with fewer data, as well as inference on how the delay mechanism varies across the different areas. The model is therefore extended to include spatial (Gaussian) random effects. Considering spatial variation where $s \in S$ denotes a spatial location or area in some spatial domain $S$, the model is now:
\begin{equation}
\label{likelihood2}
n_{t,d,s} \sim NegBin(\lambda_{t,d,s}, \phi), \qquad \lambda_{t,d,s} >0, \quad \phi>0,
\end{equation}
where $n_{t,d,s}$ is the number of occurrences in spatial location $s$ and time point $t$, reported with delay $d$ time points. In the first instance, the mean is then modelled as
\begin{equation} \label{lambda2}
\log(\lambda_{t,d,s}) =  \mu + \alpha_t + \beta_d + \gamma_{t,d} + \eta_{w(t)} + \psi_s + \beta_{d,s} + \bm{X}'_{t,d,s}\bm{\delta}
\end{equation}
where $\bm{X}'_{t,d,s}$ is now a model matrix that may also contain spatially varying covariates. The quantities $\alpha_t$ and $\beta_d$ are defined in the same way as before, but are now respectively interpreted as the overall temporal and delay evolution across space. The component $\beta_{d,s}$ captures the way in which the delay structure varies across space, while $\psi_s$ describes the overall spatial variability and dependence in the counts. The particular formulation is motivated by the application to SARI data, where the spatial region is fairly small so the temporal effects ($\alpha_t$) are a not assumed to vary with space. Given the implementation of the model in R-INLA, various possible choices exist for the specific formulation of $\beta_{d,s}$ and $\psi_s$. The space-time or space-delay interactions can range in complexity, from spatially and temporally unstructured Gaussian processes to non-separable formulations (see \cite{knorrheld2000bayesian, Blangiardoetal2013}). The spatial effect $\psi_s$ can be defined by an Intrinsic AutoRegressive (IAR) process \citep{Besag1991} if the data are counts in areal units, to allow similar temporal variation in neighbouring areas.  Equally, $\psi_s$ can be defined by a stationary Gaussian process if the data are counts in point locations, e.g. so that spatial dependence decreases exponentially with distance. In the application of the model in section \ref{sec:sari}, where space is divided in a number of administrative areas, we use the type I space-time interaction as proposed by Knorr--Held (2000) \cite{knorrheld2000bayesian}. This is a formulation where 
\begin{equation}
\beta_{d,s} \sim N( \beta_{d-1,s}, \omega^2_\beta)
\end{equation}
is an independent first order random walk for each ares $s$, and where $\psi_s=\psi_s^{IAR}+\psi_s^{ind}$ i.e. the sum of a spatially structured IAR {process:
$$
\psi_s^{IAR}|\psi_{s'\neq s}^{IAR} \sim N\left( \frac{\sum_{s'\neq s}w_{s,s'}\psi_{s'}^{IAR}}{\sum_{s'\neq s}w_{s.s'}},\frac{\sigma^2_{IAR}}{\sum_{s'\neq s}w_{s,s'}} \right)
$$
and spatially unstructured random effects $\psi_s^{ind}\sim N(0,\sigma^2_{ind})$. Here, $\sigma^2_{IAR}$ controls the strength of spatial dependence and $\sigma^2_{ind}$ is the variance of the spatially unstructured effects.}

\subsection{Nowcasting}\label{sec:nowcast}
In any given time step $T$, there are a number of occurred-but-not-yet-reported (missing) values $n_{t,d}$, $t=T-D+1,\ldots,T; d=1,\ldots,D$ (grey cells in Table \ref{table1}), as well as the marginal totals $N_{T-D+1},\ldots,N_T$. Of primary interest is of course $N_T$ which needs to be nowcast, however hindcasts of $N_{T-D+1},\ldots,N_{T-1}$ may also be of interest, especially if one wants to quantify the rate of increase or decrease in the counts.

From a Bayesian perspective, this is a prediction problem where all the missing $n_{t,d,s}$ can be estimated from the posterior predictive distribution
\begin{equation}\label{predictive}
p(n_{t,d,s}|\bm{n}) = \int_{\bm{\Theta}} p(n_{t,d}|\bm{\Theta})p(\bm{\Theta}|\bm{n}) d\bm{\Theta}
\end{equation}
where $\bm{n}$ denotes all the data used to fit the model. This cannot be solved analytically, however with samples from the posterior $p(\bm{\Theta}|\bm{n})$ one can use Monte Carlo to approximate \eqref{predictive}. In practice, for each sample from $p(\bm{\Theta}|\bm{n})$, we simulate a value from the Negative Binomial $p(n_{t,d}|\bm{\Theta})$ to obtain an approximate sample from the predictive distribution $p(n_{t,d}|\bm{n})$. Due to the autoregressive nature of the temporal and delay components, predictions are performed sequentially starting from the top right corner of the run-off triangle, i.e. $n_{T-D+1,D}$, then moving down the rows sequentially going from left to right column-wise. Once posterior predictive samples of $n_{t,d}$ are available, then it is a matter of arithmetic to obtain equivalent samples from $p(N_t)$, the marginal totals. Samples from an approximation of the joint posterior distribution $p(\bm{\Theta}|\bm{n})$ can be obtained from R-INLA using the \texttt{inla.posterior.sample()} function as also illustrated in small area estimation (e.g. \cite{Vandendijck2016spatialstats}).

Ultimately, one has to be conscious of the approximations involved in using the INLA approach at the gain of significant increase in computational speed. In the next section, we perform a comparison between the non-spatial version of the model in \eqref{likelihood} when implemented using both MCMC and R-INLA.

\section{Model application}\label{sec:application}

In this section we apply the proposed model to two situations relating to infectious disease in Brazil. The first involves correcting reporting delay for the occurrence of dengue fever in the city of Rio de Janeiro, while the other relates to correcting reporting delay for Severe Acute Respiratory Infection (SARI) across the Brazilian state of Paran\'{a}. Both implementations of the model are now being used as decision making tools by local and national authorities with an associated interface to online warning systems, infoDengue (\url{https://info.dengue.mat.br}) and infoGripe (\url{http://info.gripe.fiocruz.br}). 

\subsection{Dengue fever in Rio de Janeiro, Brazil}\label{sec:dengue}

Dengue fever is an infectious vector-borne disease that has been endemic in Brazil since 1986. The transmission of dengue is characterized by significant year-to-year variability, driven by the complex interactions between environmental factors (such as temperature and humidity), human factors (such as population immunity and mobility), and viral factors (circulating strains). Uncertainties in these interactions impair the ability to prepare for and allocate resources to reduce disease burden. In this context, continuous surveillance, fast analysis and response are key for a successful control. {In principle, dengue is meant to be reported within seven days of case identification.} However, in practice less than 50\% of the cases are reported within one week, less than 75\% within four weeks and no more than 90\% within 9 weeks. Therefore, a reasonable upper bound for the delay time $D$ is 10 weeks.

\begin{figure}[ht] 
\centering
\includegraphics[width=0.8\textwidth]{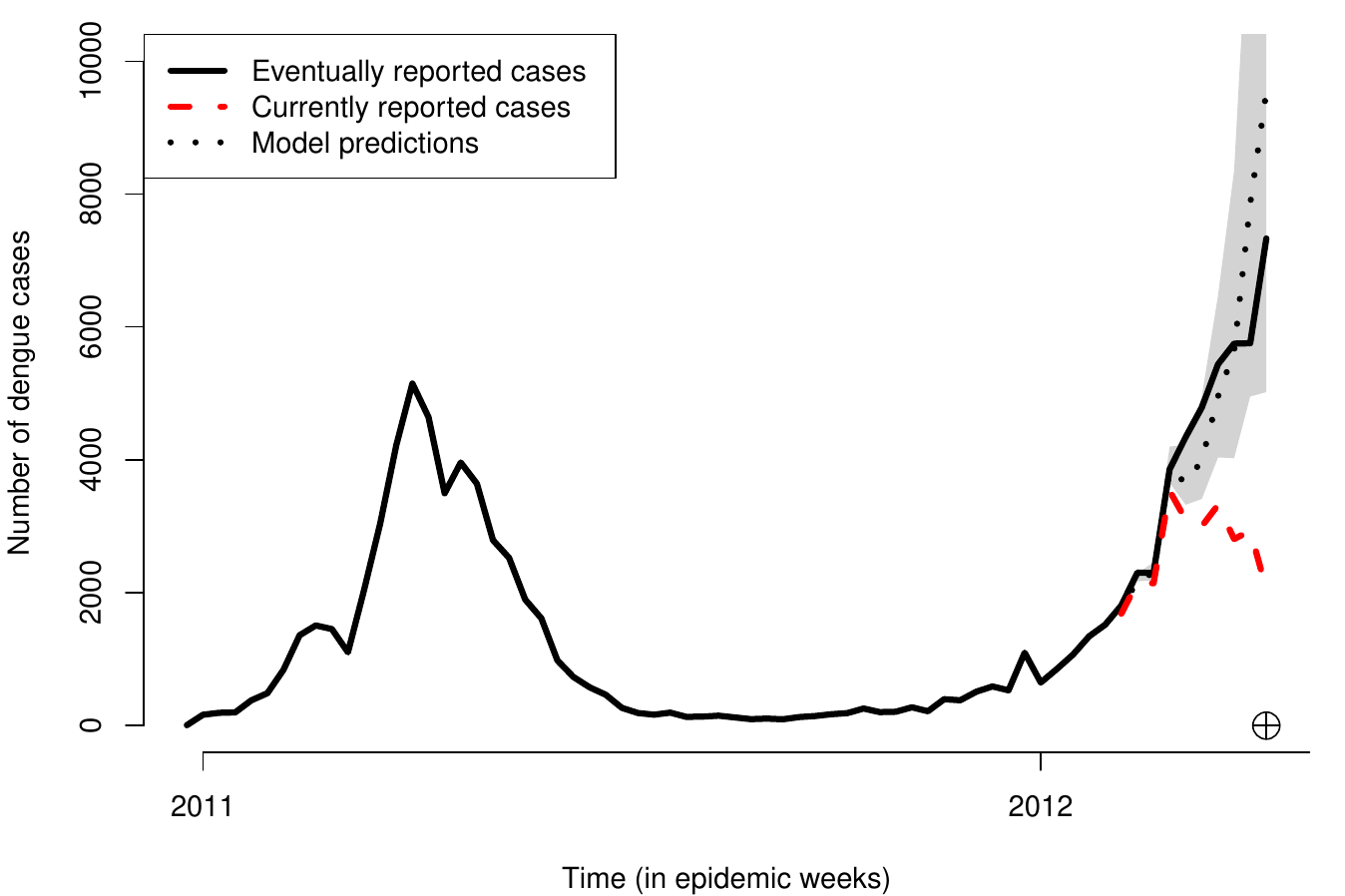}
\caption{Time series of reported dengue cases in Rio de Janeiro, from January 2011 to April 2012.  The black solid line shows the {eventually reported number of dengue cases} per week. The red dashed line shows the {currently} reported number of cases from the 6th to the 15th epidemic week of 2012 (circled cross). The black dotted line shows the model estimates for this period, along with 95\% prediction intervals in grey.} 
\label{delay.fig}
\end{figure}

Reported suspected cases of dengue, as recorded by the Brazilian Information System for Notifiable Diseases (SINAN and DENGON) were provided by the Rio de Janeiro Health Secretariat. Records include two dates: date of reporting (when doctor fills in the reporting sheet) and date of digitisation (when the sheet is fed into the system) and the former is used as a reference for $t$. Date of disease onset, although available, presented a large percentage of missing values. The importance of timeliness for decision making is easily observed in a time series plot of dengue cases in Rio de Janeiro presented in Figure \ref{delay.fig}, where {the `eventually reported' number of cases during the 2012 outbreak (black line) is considerably larger than `currently reported' number of cases reported (red line) as of the 15th epidemic week of 2012} (from the 7th to the 13th of April 2012). {Notice that if only the current number of cases is considered, a public health decision maker could potentially take wrong actions on the basis that the number of dengue cases appears to be decreasing. Also worth noting is the fact that eventually reported cases are corrected for both delays and occasionally for misclassification using laboratory confirmation tests.}

\subsubsection{Results and model checking}

The available data consists of weekly counts of the number of dengue cases in Rio for the time period January 2011 -- December 2012, along with the associated delay information. The model used for estimation is the one given by \eqref{likelihood} with $D=10$ and $\bm{X}_{t,d}=\bm{0}$ as no covariate information was available. A single run of the model was performed with time $T$  being the 15th epidemic week of 2012 (see Figure \ref{delay.fig}), meaning that $t=1,\ldots,68$ weeks and $T=68$. The model was used to correct the total number of cases $N_T$ in that particular week, but also for the 9 weeks preceding it, as shown in Figure \ref{delay.fig} (black dotted line), along with 95\% prediction intervals. The plot indicates that the predictions actually identify the fact that there is an outbreak.

\begin{figure}[ht] 
\centering
\includegraphics[width=0.6\textwidth]{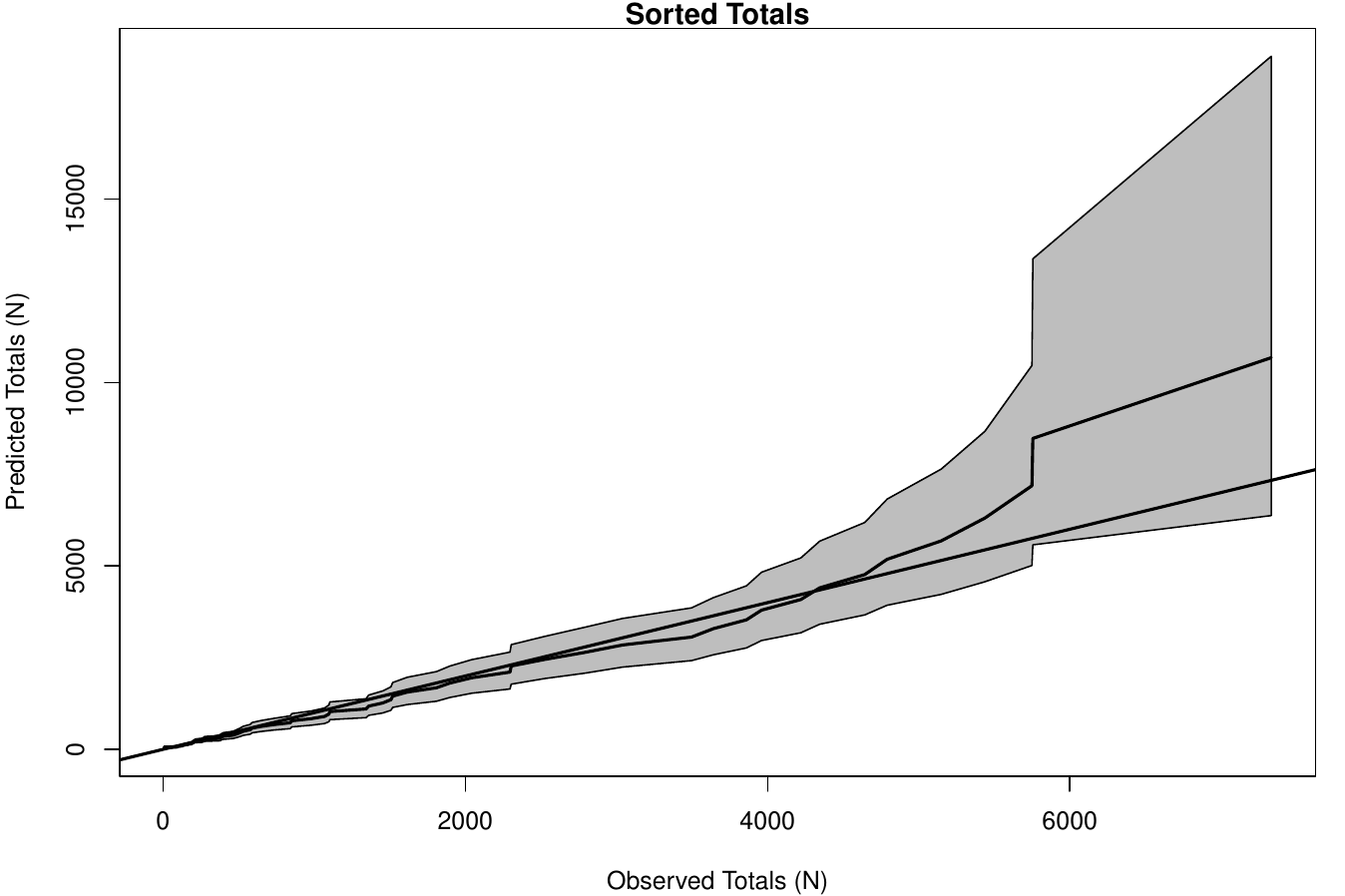}
\caption{Predicted totals plotted against the respective observed (sorted) values.} 
\label{fig:fitted}
\end{figure}

{To ensure the model provides a good fit to the data, we conduct a series of checks. First the predictive distributions of the totals $N_1,\ldots,N_{68}$ are computed from summing the respective $n_{t,d}$ over $d$. Figure \ref{fig:fitted} shows the predicted $N_t$ defined as the means of these distributions, plotted against the observed $N_t$ sorted in ascending order. The 95\% prediction intervals are also added, and the plot indicates that the model estimates capture the rank of the observed values very well, bearing in mind that 10 of the 68 values are based on data the model has not seen.}

\begin{figure}[ht] 
\centering
\includegraphics[width=0.9\textwidth]{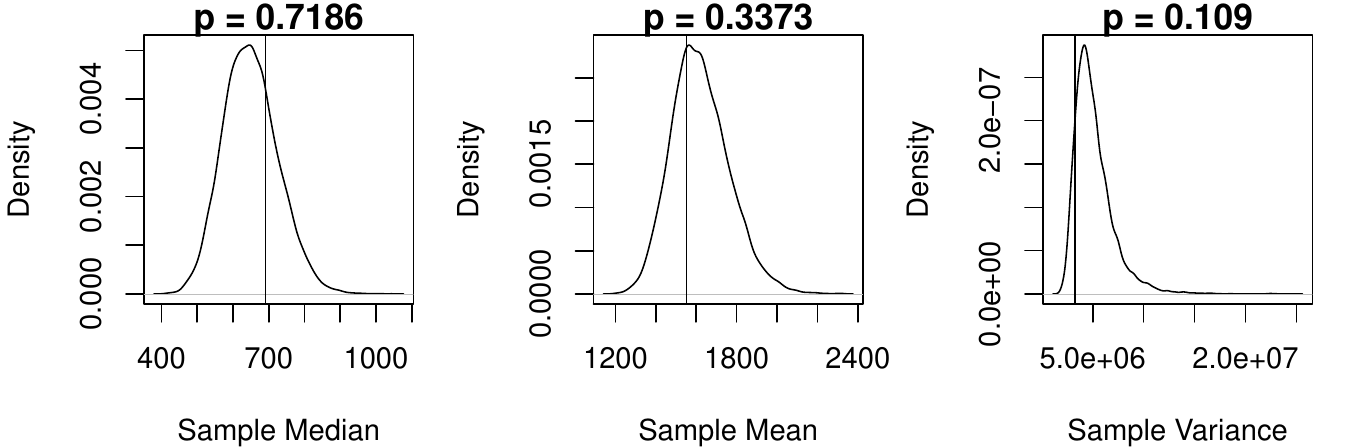}
\caption{Predictive distributions for the sample mean, median and variance of the totals $N_t$. Vertical lines indicate the observed values while, the quoted probabilities indicate the tail area of the observed values (values less than 0.025 or over 0.975 indicate the observed value is not well represented by the model).} 
\label{fig:MMV}
\end{figure}

\begin{figure}[ht] 
\centering
\includegraphics[width=0.9\textwidth]{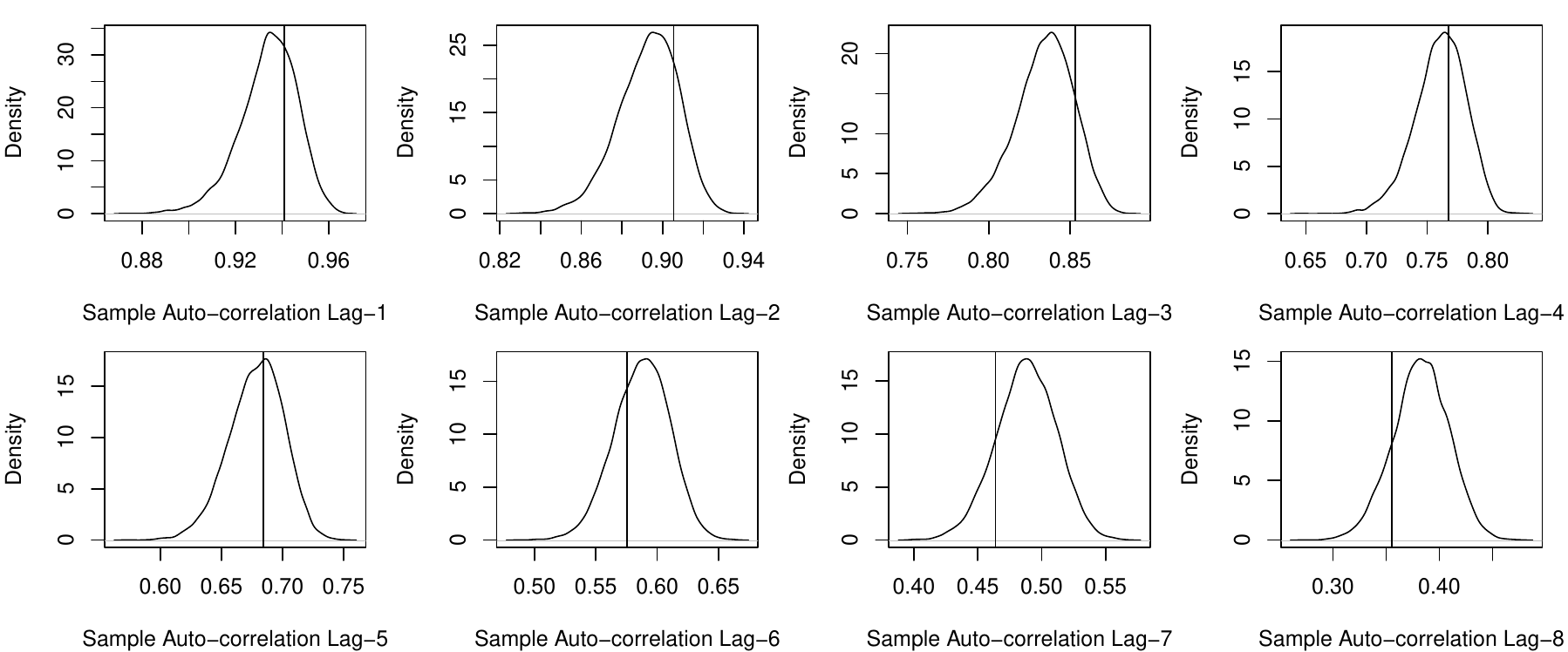}
\caption{Predictive distributions for the sample auto-correlation of the totals $N_t$, for the first 8 lags. Vertical lines indicate the observed values.} 
\label{fig:acf}
\end{figure}

{Furthermore, we look at the sample mean, median and variance of the totals $N_t$ and check how well these are captured by computing the respective posterior predictive distributions of these three statistics (from predictive samples of the totals). The plots in Figure \ref{fig:MMV} indicate that the three statistics are well-captured (i.e. are not extreme with respect to the distributions). Also, to check whether temporal dependence in $N_t$ is well captured, we consider the sample auto-correlation in the $N_t$ for the first 8 lags. Figure \ref{fig:acf} shows that these are well-captured by the model, since the none of the observed values (vertical lines) are extreme with respect to the respective predictive distributions.}

{A further aspect of the data we would like to ensure the model captures well, is the covariance, $\mbox{Cov}(n_{t,d},n_{t,d'})$ of the various columns in the data matrix (Figure \ref{table1}). To that end, we produce many replicates of the data matrix from the respective predictive distributions of $n_{t,d}$, from which we compute the predictive distribution of the sample covariance between each column. We then compute the lower tail area probability of the observed sample covariance value (as in Figure \ref{fig:MMV}). Table \ref{tbl:cov} shows these tail area probabilities, noting that only 3\% of these are extreme, i.e. smaller than 0.025 or larger than 0.975, indicating that the covariances are well captured.
}

\begin{table}[ht]
\caption{Lower tail area probabilities quantifying how well the model captured the sample covariance of each column in the data. Only upper triangular is shown as the matrix is symmetric.} \label{tbl:cov}
\centering
\begin{scriptsize}
\begin{tabular}{llllllllllll}
Delay   &    0    &  1 & 2 & 3 & 4 & 5 & 6 & 7 & 8 & 9 & 10    \\ 
\hline
0 & 0.110 &0.200 &0.164& 0.097& 0.099& 0.048& 0.129 &0.169 &0.263 &0.660 &0.660 \\
1 & & 0.164 &0.302 &0.103 &0.131 &0.104 &0.152& 0.091& 0.105& 0.296 &0.630\\
2 &&&0.146 &0.251& 0.133& 0.128& 0.114& 0.057& 0.044& 0.256 &0.670\\
3 &&&& 0.212 &0.267 &0.043& 0.036 &0.036 &0.022& 0.059 &0.440 \\
4 &&&&& 0.223& 0.130& 0.117& 0.050& 0.063& 0.020 &0.432\\
5 &&&&&& 0.138& 0.170& 0.025 &0.028 &0.038 &0.199\\
6 &&&&&&& 0.166 &0.114 &0.061& 0.169& 0.260\\
7 &&&&&&&& 0.092 &0.110 &0.238 &0.192\\
8 &&&&&&&&& 0.122 &0.430 &0.068 \\
9 &&&&&&&&&& 0.673 & 0.660 \\
10 &&&&&&&&&&& 0.811\\
\hline
\end{tabular}
\end{scriptsize}
\end{table}

\subsubsection{Sensitivity analysis of using INLA for prediction}
As discussed in section \ref{sec:nowcast}, predictions from R-INLA are approximate and it therefore makes sense to assess the effect of the approximations. For the same data depicted in Figure \ref{delay.fig}, the model used in the previous subsection is implemented using MCMC. More specifically, the R package NIMBLE \citep{nimble} was used, which uses a combination of Gibbs sampling and the Metropolis-Hastings algorithm. Three chains were run for 2.5 million iterations each, a burn-in of 2 million and thinning of 10 (totalling 150000 samples), to ensure convergence and good mixing. {Convergence was assessed by visual inspection of trace plots, and by computing the Potential Scale Reduction Factor (PSRF,\cite{Brooks}). This compares the variance between the MCMC chains to the variance within the chains. A PSRF of 1 is obtained when the two variances are the same, so starting the chains from different initial values and obtaining a PSRF close to 1 (typically taken to be less than 1.05) gives a good indication of convergence to the posterior distribution. Out of the 884 random effects and hyperparameters in the model, the maximum PSRF was 1.034 indicating all have converged according to this measure. In addition, we compute the effective number of samples per hyperparameter, by accounting for the autocorrelation in the chains. The smallest effective number of samples was 1420, ensuring that there are enough independent samples for conducting iference based on Monte Carlo.}

\begin{figure}[ht] 
\centering
\includegraphics[width=0.8\textwidth]{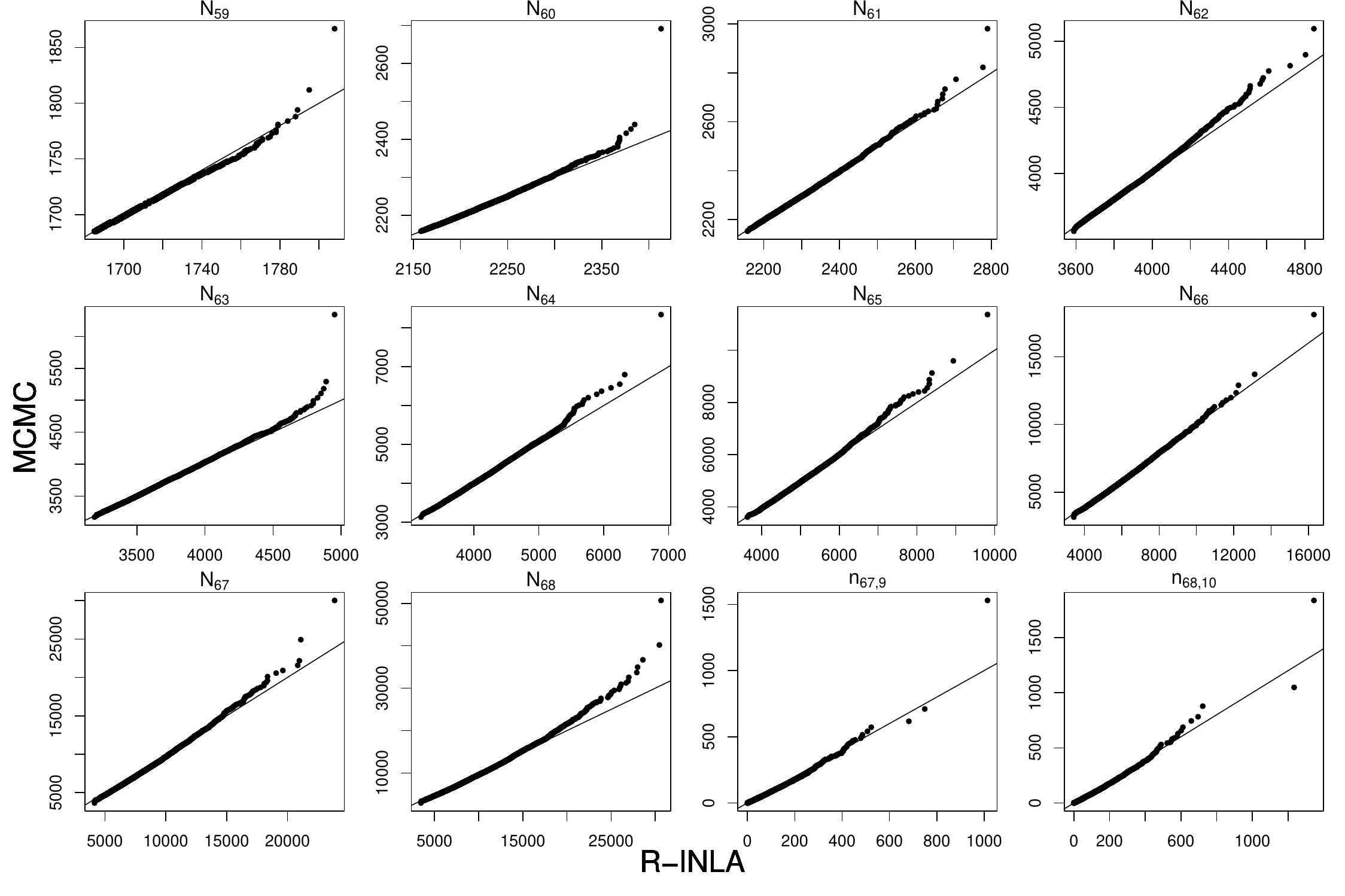}
\caption{Q-Q plots comparing R-INLA and MCMC samples from the predictive distribution of the total counts $N_t$, for $t=59,\ldots,68$ where $T=68$ is the 15th epidemic week of 2012.} 
\label{fig:compareN}
\end{figure}

R-INLA and MCMC samples from the predictive distributions of the 10 unknown total counts $N_t$ are compared using Q-Q plots shown in Figure \ref{fig:compareN}. To also compare some of the individual counts $n_{t,d}$, the last two panels on the bottom right of Figure \ref{fig:compareN} compare samples from the predictive distributions of $n_{T,D-1}$ and $n_{T,D}$, i.e. the last two entries of the ``$T$'' row in Table \ref{table1}. Overall, the predictive distributions match well with the exception that R-INLA samples sometimes tend to slightly under-estimate the upper tail. This conclusion is representative of further comparisons across other weeks not shown for conciseness. We feel that this is a reasonable compromise to the gain in computational speed with the R-INLA model taking a matter of seconds compared to hours of MCMC (per chain).

\subsubsection{Estimates and rolling predictions}
\begin{figure}[ht] 
\centering
\includegraphics[width=0.9\textwidth]{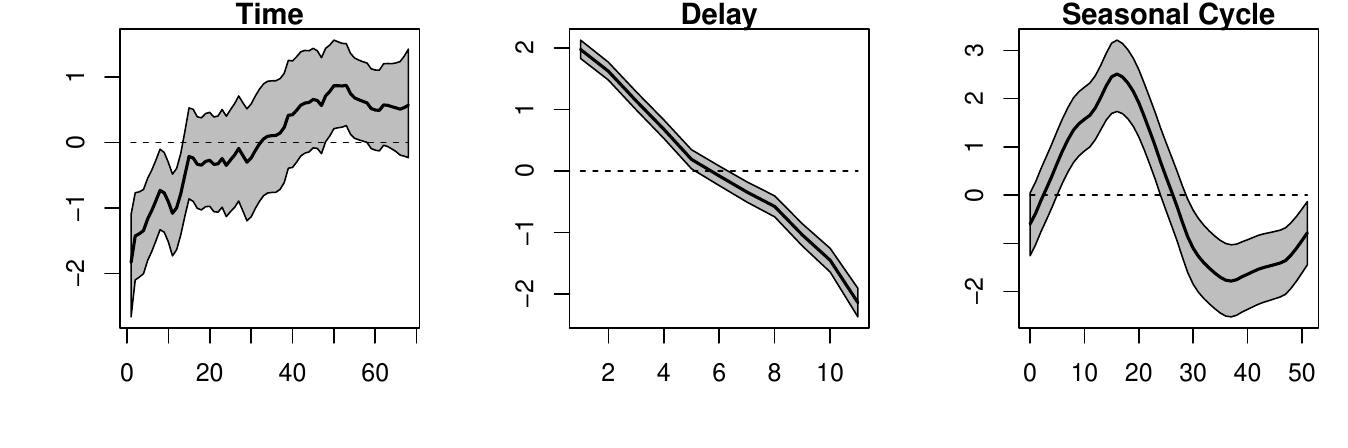}
\caption{Estimates of the overall temporal variation in the counts (left), the overall delay structure (middle) and the seasonal variability (right).} 
\label{fig:RE}
\end{figure}

{Figure \ref{fig:RE} shows estimates of three components, namely $\alpha_t$ the overall temporal evolution in the counts, $\beta_d$ the delay structure, and $\eta_{w(t)}$ the seasonal variability. The overall temporal effect is increasing at first but then plateaus, perhaps reflecting an increase in the susceptible population. The delay structure is almost linear and decreasing, as would be expected---the more time goes by, the more cases are being reported. There is also a strong seasonal component capturing an increase in the early part of the year and a decrease later on.}

\begin{figure}[ht] 
\centering
\includegraphics[width=\textwidth]{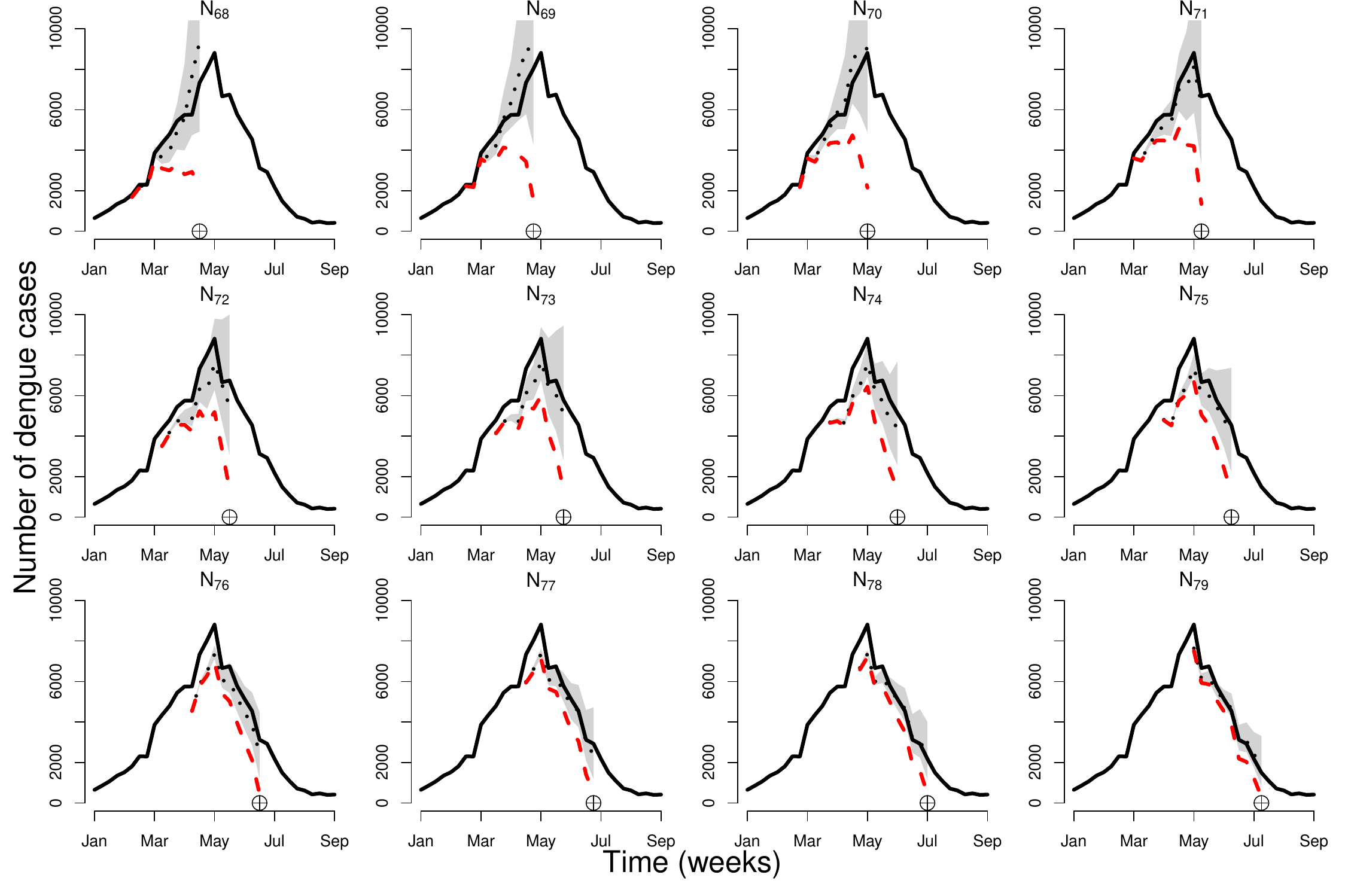}
\caption{Time series of dengue cases in Rio de Janeiro for 12 epidemic weeks starting from the 15th epidemic week of 2012 on the top left {($T=68$)}. The black line shows the {eventually reported} number of cases; the red dashed line shows the number of {currently reported} cases; while the black {dotted} line shows model predictions {(of the eventually reported number of cases)} along with 95\% prediction intervals. The circled cross symbol indicates {the epidemic week $T=68,69,\ldots,79$.}} 
\label{fig:predsN}
\end{figure}

Furthermore, Figure \ref{fig:predsN} shows weekly rolling predictions, starting from the 15th going to the 26th epidemic week of 2012. This period was chosen specifically to test the ability of the model to capture the outbreak as well as the relatively sharp decline, {in the eventually reported number of cases (black line)}. From Figure \ref{fig:predsN} it is evident that the model (black dotted line) captures both the increase and decrease in the {eventually reported} number of cases, despite a sharp decrease in the week that the peak occurs (top right panel). It is important to note that most of the {eventually reported} counts are within the 95\% prediction intervals, particularly for time $T$ (indicated by the circled cross), which is the most important value.

Predictions from the particular model implementation presented here are currently being used to inform a disease warning system in 790 Brazilian municipalities from six Brazilian states, Rio de Janeiro, Paran{\'a}, Esp{\'{\i}}rito Santo, Cear\'a, Minas Gerais, and S\~ao Paulo. The warning system also uses Twitter feeds and weather information but the primary source of information for predicting both dengue and chikungunya cases (another mosquito-borne disease) are the posterior predictive means from the model presented here. Note that the model is being fitted independently in each region, but in the next section an application to SARI involves a spatial structure.

\subsection{Severe acute respiratory infection (SARI) in Paran\'{a}, Brazil} \label{sec:sari} 

The lack of a baseline for detecting changes in disease severity during the 2009 H1N1 Influenza pandemic led the World Health Organization (WHO) to standardize, in 2011, a definition for the notification of SARI worldwide. SARI is defined as an acute respiratory infection with onset within the past 10 days, and a history of fever or measured fever above $38^oC$, coughing, and requiring hospitalization \citep{fitzner2018revision}. The goals of SARI surveillance include, among other things, to determine the seasonal patterns of respiratory virus circulation, to detect the emergence of high pathogenicity influenza viruses and to provide timely information to guide prevention policies.

In general, weekly reports of SARI activity are sent from hospitals to the local (municipality) health authorities, then aggregated at the state and national levels, and eventually sent to the WHO. 
Besides total case counts of SARI, laboratory tests are carried out to identify the etiological agent associated and provide specific diagnosis. This procedure also enables stratification of the number of SARI cases per type of virus. Each one of those steps introduce delays in the information available for epidemiological situation rooms, created at local or global levels. For influenza, having a precise estimate of SARI activity in a timely manner is fundamental to update the indicators of activity upon which decisions are made. 

In 2009, the Brazilian state of Paran\'a was heavily affected by the H1N1 epidemic, accounting for 52\% of all reported cases {in Brazil} and an incidence rate of SARI at least four times greater than the other states \citep{codecco2012epidemic}. Brazil implemented the national SARI surveillance in 2009  and since then Paran\'a remains among the states with largest attack rates. Paran\'a is an important point of entry from Argentina and Paraguay into Brazil, has an intense touristic activity, and has an important poultry industry (type of landscape at risk of emergence of new influenza viruses). As such, implementing a nowcasting SARI surveillance has strong practical implications.

\subsection{Data}
The data consists of SARI reports extracted from  the Brazilian Information System for Notifiable Diseases (SINAN) starting on 01 January 2016 and ending on 2nd April 2017 (66 weeks) for the state of Paran\'a. The state is divided into 399 municipalities and each municipality belongs to one of 22 health regions. The available data are aggregated at the health region level. The goal is to use the proposed model in order to correct reporting delay across the health regions, taking into account spatial variability in the delay mechanism and the disease process, as well as allow for spatial dependence in neighbouring health regions. In particular we consider the model described in section \ref{sec:spatial}, namely:
\begin{eqnarray} \label{SARImodel}
n_{t,d,s} &\sim& NegBin(\lambda_{t,d,s}, \phi) \nonumber \\
\log(\lambda_{t,d,s}) & = & \mu + \alpha_t + \beta_{d}  + \gamma_{t,d} + \beta_{d,s} + \psi_s^{IAR}+\psi_s^{ind} \label{Mean}
\end{eqnarray}
with $t=1,\ldots,66$ (weeks), $d=0,\ldots,10$ (delay weeks) and $s=1,\ldots,22$ (health regions). 
The quantities $\alpha_t$, $\beta_d$ and $\gamma_{t,d}$ are defined as before, while $\beta_{d,s}\sim N(\beta_{d-1,s},\omega^2_\beta)$ allow for unstructured spatio-delay variability. {Note that more complex spatio-temporal structures are possible, see for instance \cite{Bauer2016} who use a penalized splines to define space-time interaction, estimated using R-INLA.} Lastly, $\psi^{ind}_s\sim N(0,\sigma^2_{\psi})$ captures spatially unstructured variability while $\psi^{IAR}_s$ \citep{Besag1991} is spatially structured according to an IAR process with a neighbouring structure defined by a $22 \times 22$ adjacency matrix $\mathbf{W}$, where $w_{i,j} = 1$ if the health region $i$ is an administrative neighbour of health region $j$, and $w_{i,j} = 0$ otherwise.

{In model \eqref{SARImodel}, we assume that the delay structure varies with health region, through $\beta_{d,s}$, while the overall temporal evolution of the disease counts $\alpha_t$ is the same across the regions. This is because the state is fairly small and we would not expect the disease transmission to vary considerably across space. Similarly, the interaction term $\gamma_{t,d}$ is spatially constant. The term $\psi_s^{IAR}+\psi_s^{ind}$ captures overall similarity in disease counts across the health regions, however it also allows for some regions to be different (on average) if there is such evidence in the data.}

\subsubsection{Results}

\begin{figure}[ht] 
\centering
\begin{tabular}{cc}
\includegraphics[width=0.45\textwidth]{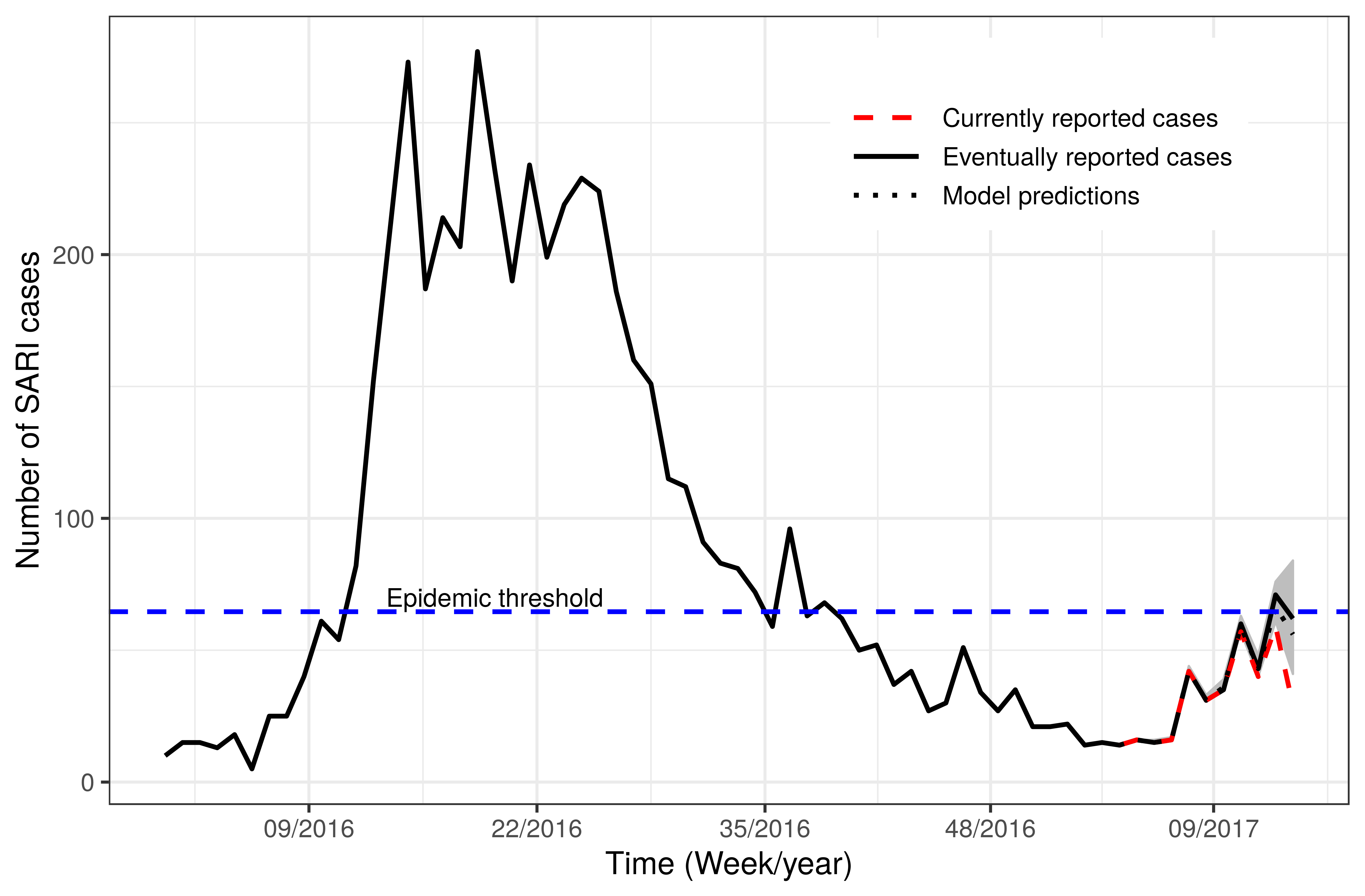}  & \includegraphics[width=0.45\textwidth]{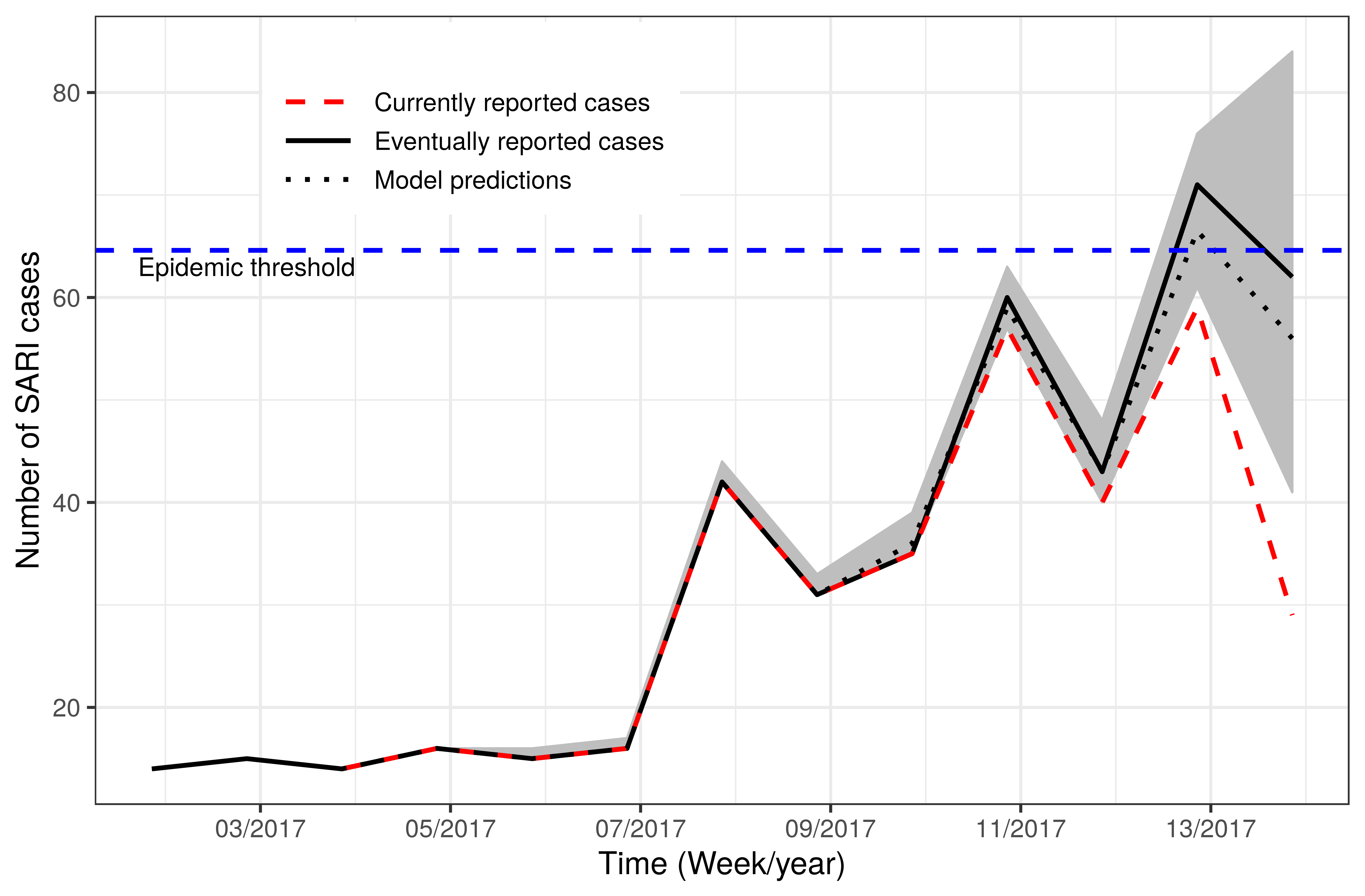} \\ 
(a) & (b) \\ 
\end{tabular} 
\caption{Time series of SARI cases reported in the whole of Paran\'a state.  The black solid line shows the true number of SARI cases, per week. The red dashed line shows the number of cases that were reported at the 14th epidemic week of 2017. The black dotted line shows the model estimates along with 95\% prediction intervals. Figures (a) and (b) differs only on the time scale, where (a) starts from January 2016 whereas (b) starts from January 2017.} 
\label{sari.fig}
\end{figure}

{Figure \ref{sari.fig} shows the weekly time series of the eventually reported SARI cases in Paran\'a from the first epidemic week of 2016 to the 14th epidemic week of 2017 (solid black line). The plot also shows the currently reported number of SARI cases for the last 10 weeks, up to and including the 14th epidemic week ending on 2nd April 2017 (dashed red line). Finally, the plot also depicts the estimated mean of the corresponding predictive distribution from model \eqref{SARImodel}, along with 95\% prediction intervals (doted black line and shaded region). The model is able to capture the increasing trend of the disease counts and the predictions are much closer to the true value compared to the currently reported counts (which actually indicate a decline).}

{At epidemic week 14 during the 2017 SARI season in Brazil, the present nowcasting strategy was able to correctly detect that the SARI activity in the state of Paran\'a likely reached a historically defined pre-epidemic level \citep{vega2013influenza}. The currently reported number of cases in weeks 13 and 14, $n_{14,0}=29$ and $n_{13,0} + n_{13,1} = 59$, were both below the epidemic threshold of 64.6 cases 
(horizontal blue dashed line in Figure \ref{sari.fig}), and it was only by the end of week 16 that the currently reported number cases of week 13 went above the threshold. In other words, the model was able to detect that epidemic activity started effectively one week after it did, while it took three weeks for the official data to detect it. In practice, this means that our system was able to detect the qualitative transition two weeks earlier, which could have been used by public health authorities to trigger mitigation strategies at the population and health practitioner level.}


\begin{figure}[ht] 
\centering
\begin{tabular}{cc}
\includegraphics[width=0.4\textwidth]{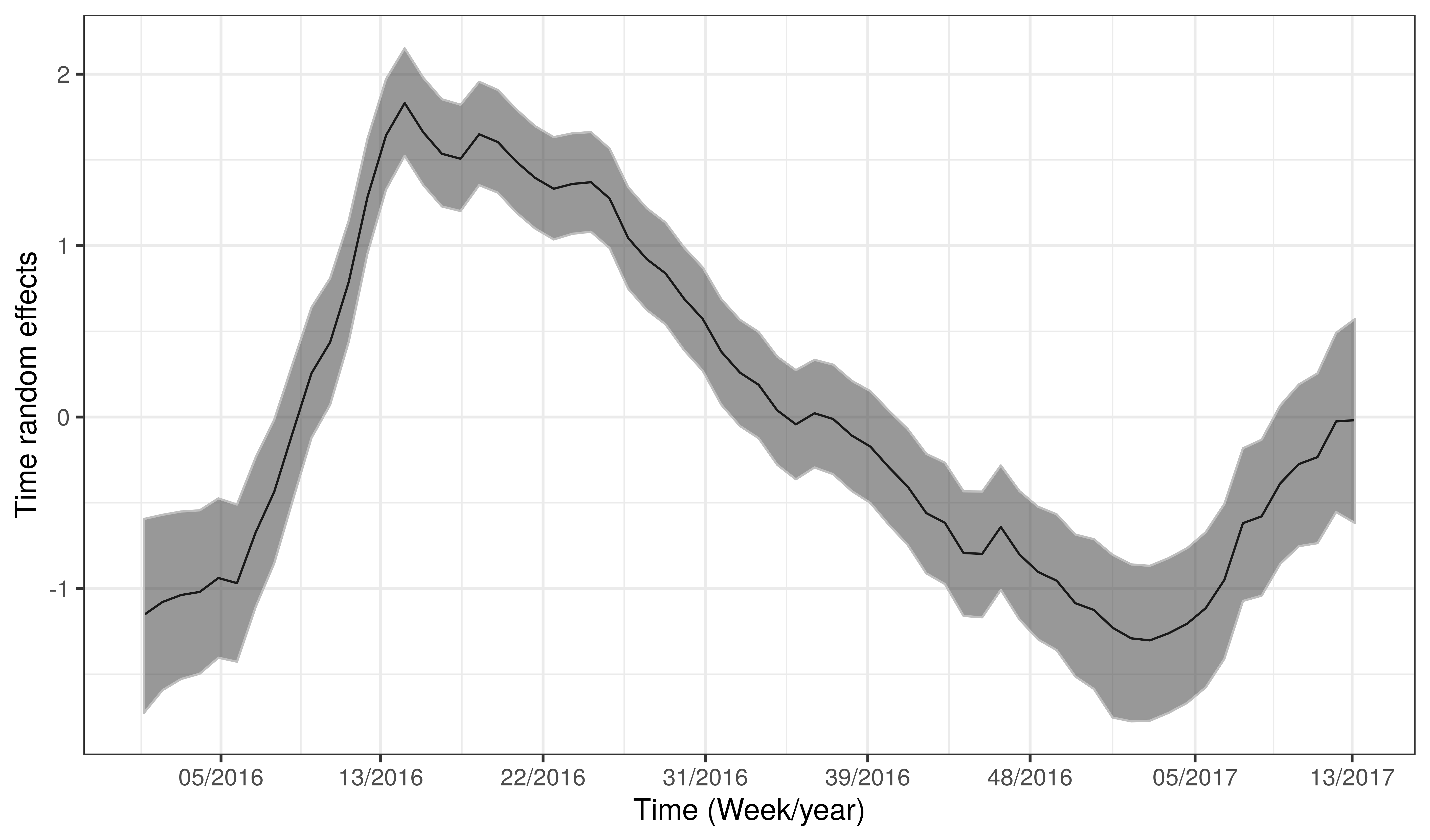} & \includegraphics[width=0.4\textwidth]{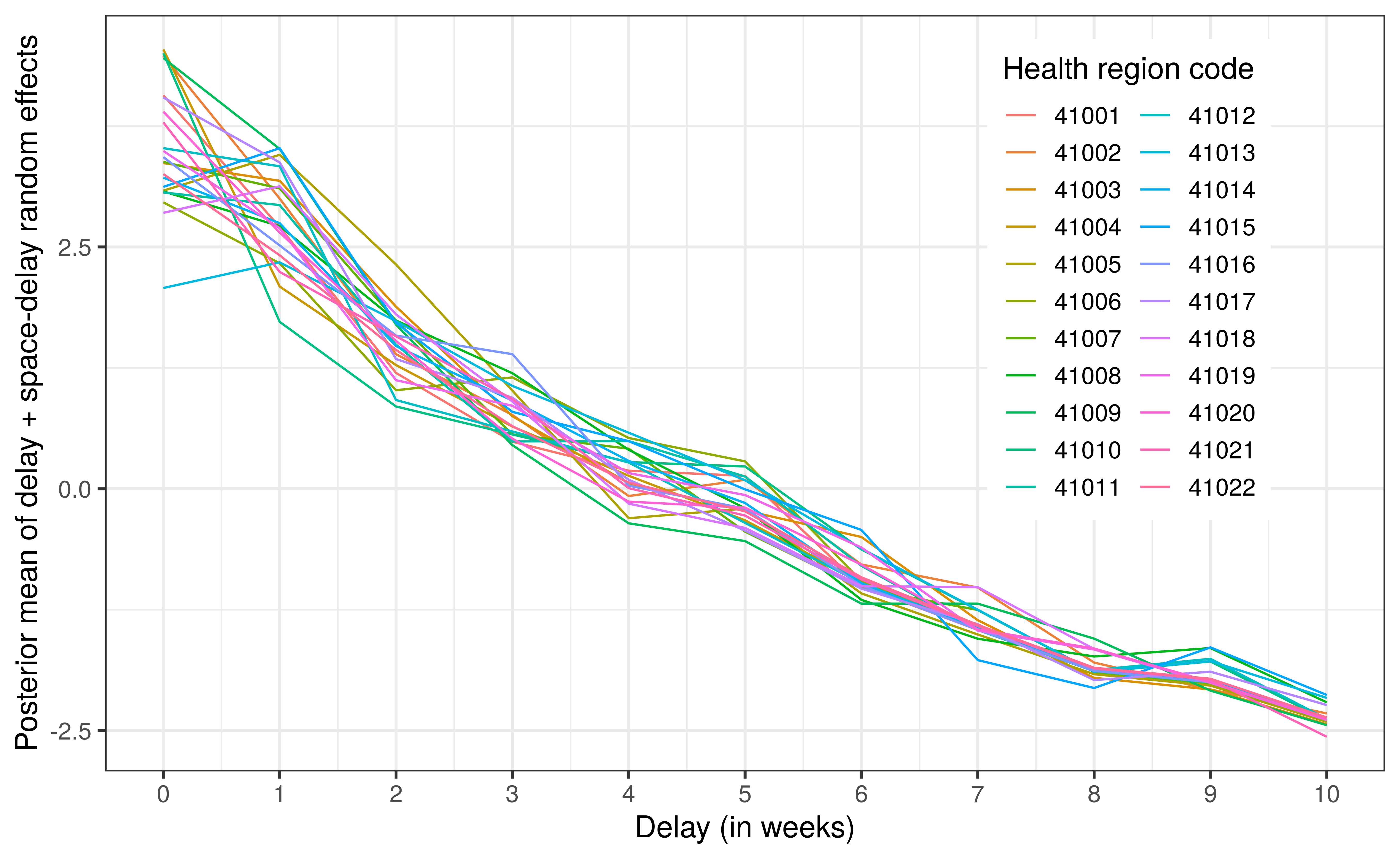} \\ 
(a) & (b) \\ 
\includegraphics[width=0.4\textwidth]{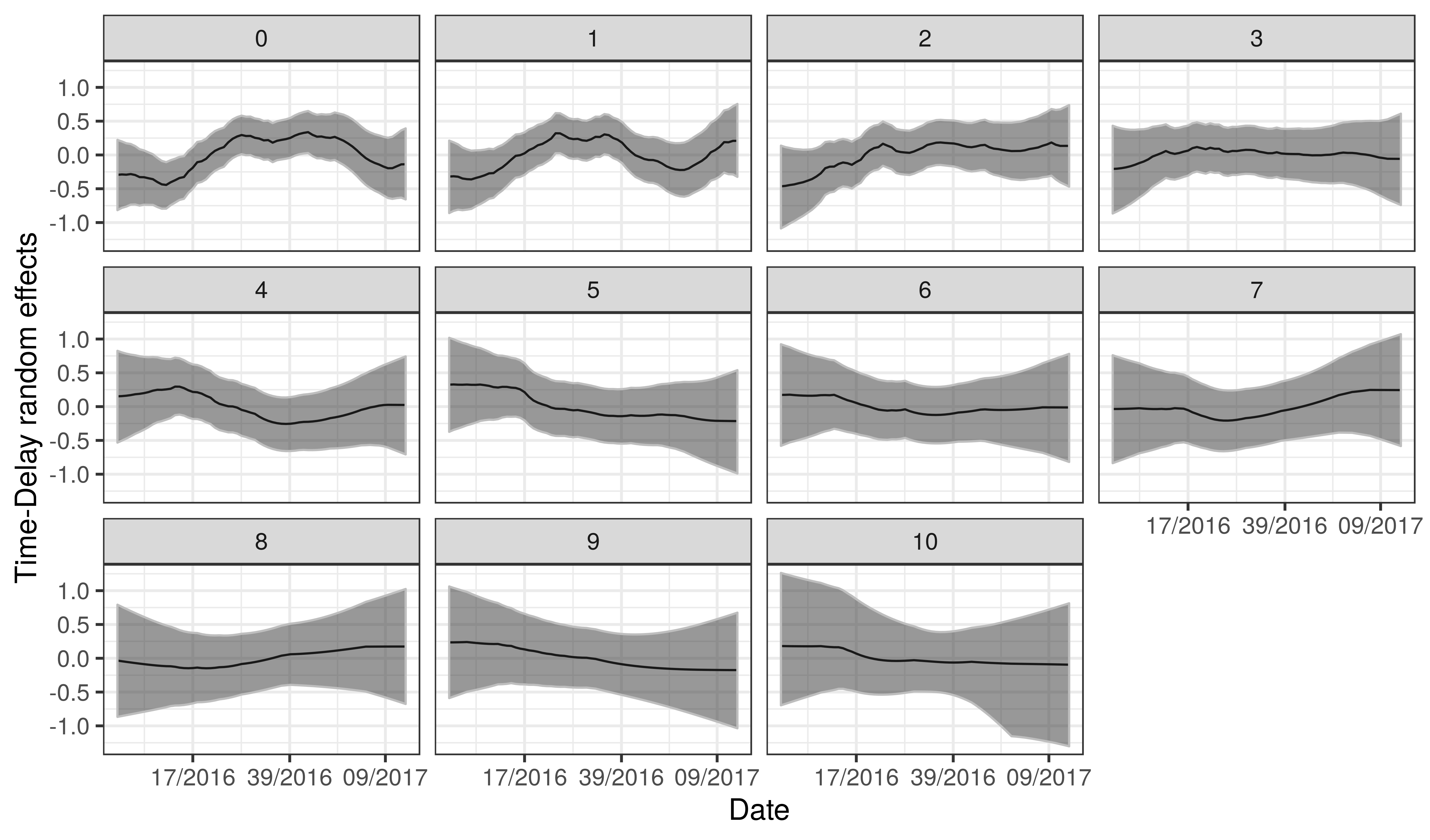} & \includegraphics[width=0.4\textwidth]{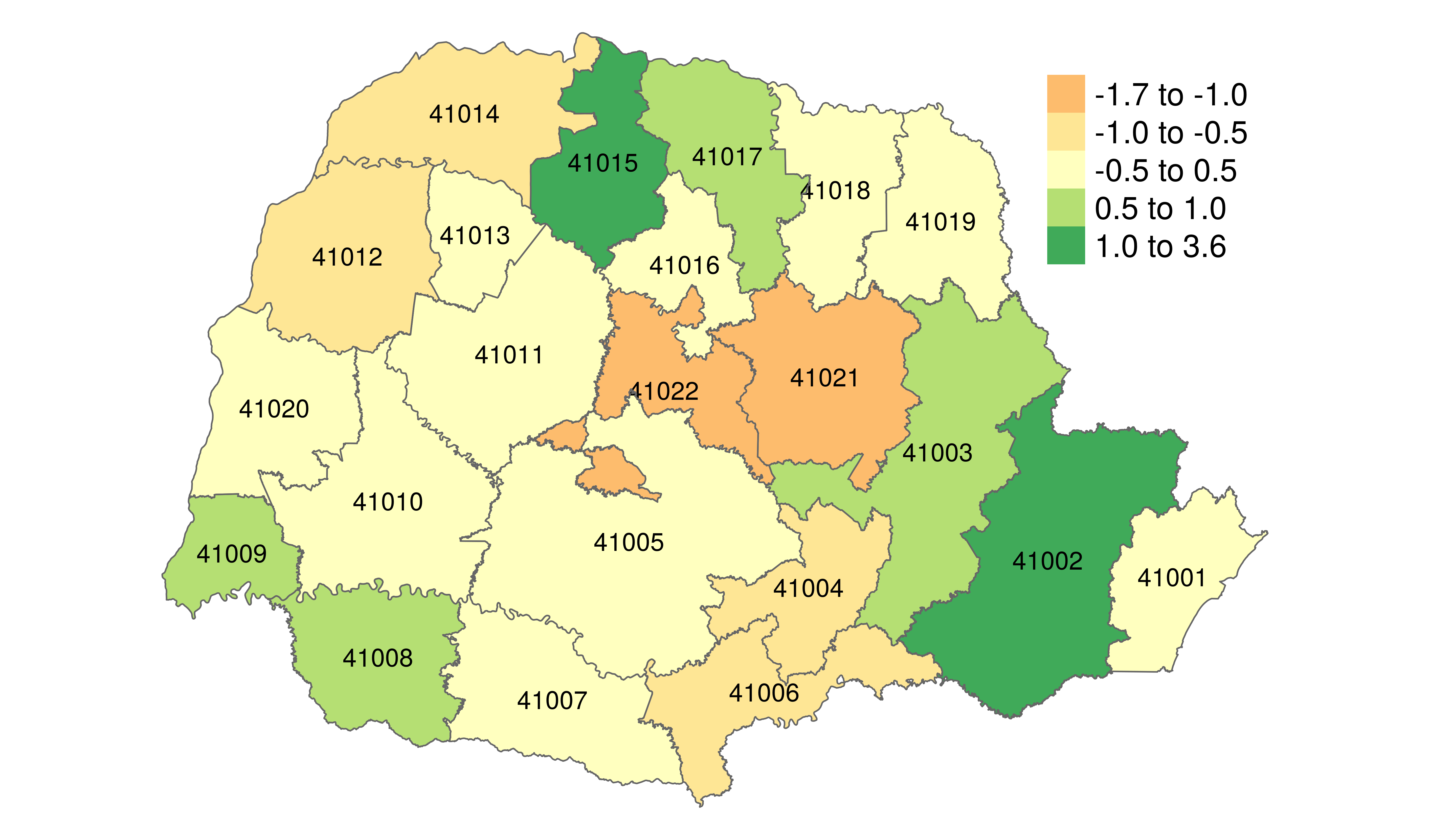} \\ 
(c) & (d) \\ 
\end{tabular} 
\caption{Estimates of the various random effects: 
(a) Posterior mean with 95\% credible intervals for time random effects, $\alpha_t$; 
(b) Posterior mean of the space-delay random effects, $\beta_d + \beta_{d,s}$ by health regions;
(c) Posterior mean of the time-delay random effects, $\gamma_{t,d}$ by delayed weeks; 
(d) Posterior mean of the spatial random effects, $\psi_s$.} 
\label{sariRE.fig}
\end{figure}

{Figure \ref{sariRE.fig} (a) shows the estimate of the temporal evolution $\alpha_t$. SARI is very sensitive to weather variations, and the state of Paran\'a which is located in the south of Brazil has well defined seasons with spring and winter being the seasons associated with the majority of SARI reports. This is reflected in the estimate of $\alpha_t$ (posterior mean and 95\% credible intervals). Estimates of the the delay mechanism, which is different across different health regions ($\beta_{d,s}$), are shown in Figure \ref{sariRE.fig} (b). On average, the mean reporting count decreases with delay (in weeks), however there is considerable variability across the health regions, particularly during the first two weeks. This eflects the fact that delay is likely related to several factors such as the region infrastructure which varies considerably in space.} 

\begin{figure}[ht] 
\centering
\includegraphics[width=0.6\textwidth]{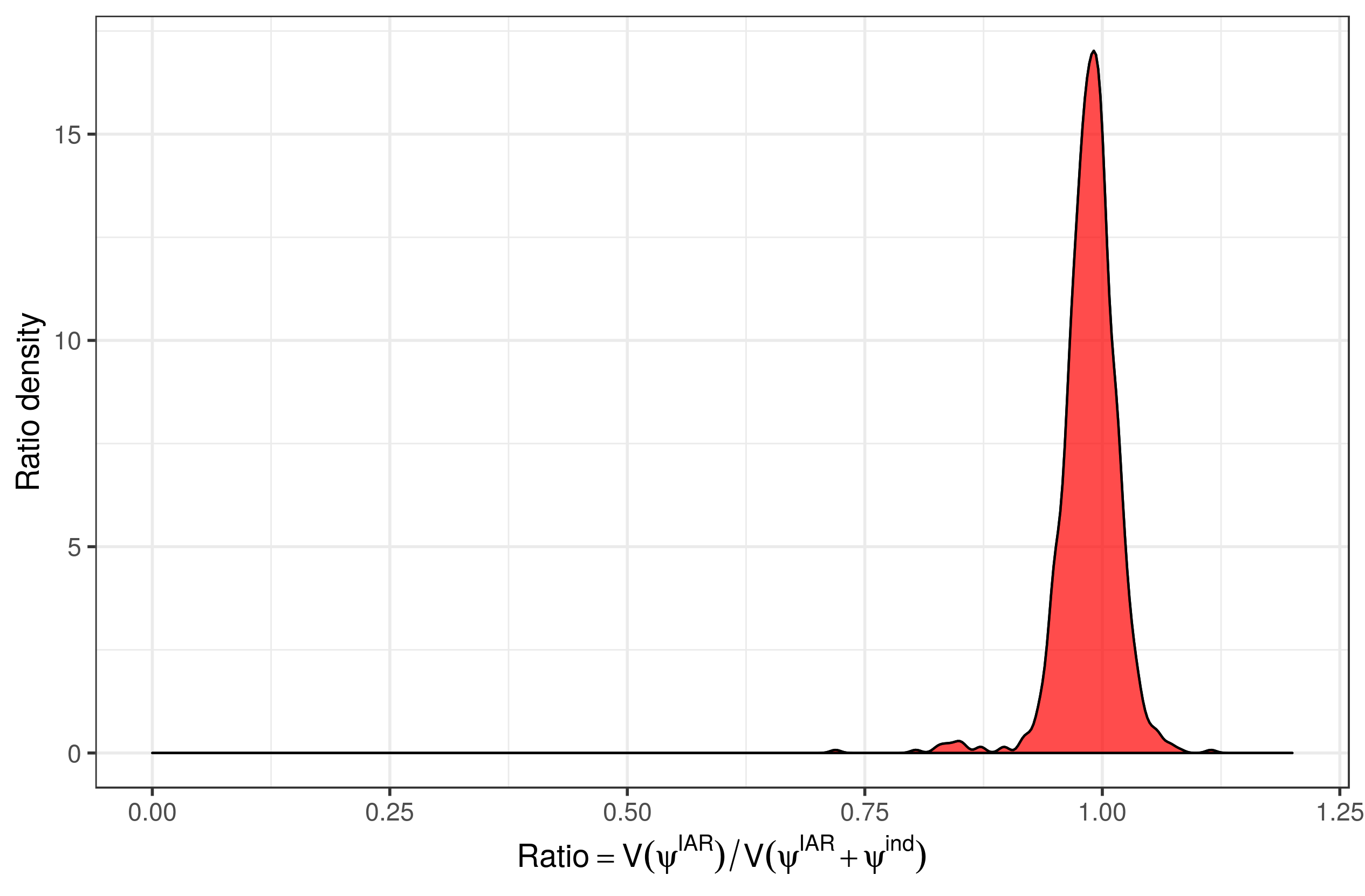}
\caption{Posterior distribution of $R$.} 
\label{fig:R}
\end{figure}

{Estimates of the time-delay interaction term $\gamma_{t,d}$ are shown in Figure \ref{sariRE.fig} (c). The plots show that the temporal evolution for $d=0$ (no delay) and $d=1$ (1 week delay) is negative in the first quarter of each year, suggesting that possible awareness of the SARI epidemic leading to faster notifications when a case is known. Furthermore, Figure \ref{sariRE.fig} (d) shows the estimate of the overall spatial variability term $\psi_s^{IAR}+\psi_s^{ind}$. This indicates some variability in the number of SARI reports across the regions, but also similarity in neighbouring regions. This is probably reflecting unobserved factors relating to the susceptible population (including population size). In order to assess whether spatial correlation was adequately captured, we consider the measure 
$$
R=\frac{\mbox{var}\left(\psi_s^{IAR}\right)}{\mbox{var}\left(\psi_s^{IAR}+\psi_s^{ind}\right)}.
$$
This quantities the contribution of the structured random effect $\psi_s^{IAR}$ to the total variance of the spatial effect $\psi_s^{IAR}+\psi_s^{ind}$. Values close to zero indicate there is not much spatial correlation while values around 0.5 indicate roughly equal contribution for structured and unstructured spatial effects. Higher values (which can be greater than 1 due to possible non-zero correlation between $\psi_s^{IAR}$ and $\psi_s^{ind}$, indicate the structured random effects are capturing most of the variability. Figure \ref{fig:R} shows a plots of the posterior distribution of $R$, which is centred at 1, indicating that the $\psi_s^{IAR}$ explains most of the variance with minimal contribution from $\psi_s^{ind}$. Had the structured random effect not be capturing spatial correlation adequately, we would expect more contribution from the unstructured effect (which can potentially compensate).}

{The R code and data for reproducing the dengue and SARI analyses are available on the aforementioned GitHub webpage: \url{https://github.com/lsbastos/Delay}.}

\section{Discussion}\label{sec:discussion}
We have presented a general modelling framework and implementation method, to flexibly model reporting delays that can in principle be applied to any disease. In fact, the proposed framework can be applied to any reporting system for which the data are described by a run-off triangle given in Figure \ref{table1}. The model was illustrated using dengue data from Rio de Janeiro and also SARI data in the state of Paran\'a in Brazil.

The two case studies, dengue and SARI, have demonstrated that the framework has desired the flexibility and complexity. In the application to dengue, a model with a dependency structure in both time and delay was utilised, while in the case of SARI data, spatial variability and dependence was assumed in order to borrow information across the spatial units and to allow for the (arbitrary) division of the data in health regions. Although none of the models included any covariates, this is a fairly trivial task in the proposed R-INLA implementation.

The implementation of the models in the Bayesian framework is extremely fast since we make use of the Laplace approximation (INLA) to compute samples from the (marginal) posteriors. In fact, the model fitted to the dengue data is currently being used to nowcast dengue cases for use in a warning system in Brazil called Info-Dengue, \url{https://info.dengue.mat.br/}. Furthermore, nowcasts from the same model are being directly used to produce warnings for influenza and SARI across the whole of Brazil by the Ministry of Health, where, for instance, the 2017 SARI outbreak in Paran\'a state was anticipated 2 weeks earlier using our proposed method. Accurate estimates of the number of disease cases are of utmost importance to avoid misclassification, e.g., failing to issue a high incidence alert. Therefore, this general method can greatly help warning analysts in surveillance systems to making well-informed decisions. Furthermore, the availability of samples from the predictive distribution of the counts implies that the predictions from the proposed models can be readily utilised in a decision theoretic framework for issuing warnings (e.g. as in \cite{Economou2016}).

\section*{Acknowledgements}
The authors would like to thank Marilia Carvalho for her support and comments. LSB, TE and TB were partially funded by \textit{Coordena\c{c}\~ao de Aperfei\c{c}oamento de Pessoal de N\'{\i}vel Superior (Capes)} grant number: 88881.068124/2014‐01.

\section*{Conflict of interest}
All contributing authors declare no conflicts of interest.


\bibliography{delaysim.bib}


\end{document}